\font\tenit=cmti10
 1
 1
 1

\documentstyle[aps,epsf]{revtex}
\draft
\def\ee{\end{equation}}
\def\be{\begin{equation}}
\title{ Phase transitions and critical behaviour in one-dimensional
nonequilibrium kinetic Ising models with branching annihilating
random walk of kinks.}
\author{\sl Nora Menyh\'{a}rd\\ {\tenit  Research Institute for Solid State
Physics,
 H-1525 Budapest,P.O.Box 49, Hungary}}
\author{\sl G\'eza \'Odor\\ {\tenit Research Institute for Materials Science,
H-1525 Budapest, P.O.Box 49, Hungary}}
\begin{document}
\maketitle

\begin{abstract}
One-dimensional  non-equilibrium kinetic Ising models evolving under the 
competing effect
of spin flips at zero temperature and nearest neighbour spin exchanges
exhibiting a parity-conserving (PC) phase transition on the level of
kinks are now  further investigated, 
numerically, from the point of view of the underlying spin system.
 Critical exponents characterising its statics and dynamics
 are reported. 
It is found that
  the influence 
 of the PC transition on the critical exponents of the
spins  is strong and the 
origin of drastic changes 
as compared to the Glauber-Ising case
can be traced back to the hyperscaling law stemming from directed percolation.
Effect of an external magnetic field, leading to directed percolation type
behaviour on the level of kinks, 
is also studied, mainly via the generalized mean field approximation.
\end{abstract}

\section{ Introduction }

Nonequilibrium phase transitions have attracted great interest lately.
A variety of sytems studied seem to belong to the universality
class of directed percolation (DP) \cite{grat,car,Jan1,grdp}
The DP universality class is very robust, among others the transitions in
branching
annihilating random walk (BARW) models with an odd number of
offsprings in the process $A\rightarrow A + nA$ belong to it \cite{taka}. 
Numerical studies by Grassberger et al. \cite{gra84,gra89} of 
probabilistic cellular automata
models in one dimension involving the processes $k\rightarrow 3k$ and
$2k\rightarrow 0$ ($k$ stands for kink) have revealed , however, a
new universality class of dynamic phase transitions. Both time-dependent
and steady-state simulations have resulted non-DP values for the
relevant critical exponents. This so called 
parity-conserving (PC) phase transition
has been found since in a variety of models. The $n=4$ BARW model
has been studied in the greatest  detail and accuracy by Jensen \cite{jen94}.
In a previous paper \cite{men94} one of the the authors has introduced 
a family of nonequilibrium
kinetic Ising models (NEKIM) showing the same phenomenon, while quite 
recently, with an appropriate modification of the original
BARW model (suggested already in ref. \cite{men94}), the $n=2$ BARW
model has been shown to exhibit the PC transition as well \cite{dani}.
The two-component interacting monomer-dimer model introduced by
Kim and Park \cite {kim94,park,parkh} 
represents a more
complex sytem with a PC-type phase transition.

 Formerly,
combinations of Glauber \cite{gla63} and Kawasaki \cite{kaw72} kinetics
were introduced with the aim of investigating temperature-driven
nonequilibrium phase transitions \cite{droz,rac94}. In ref. \cite{men94}
,however, spin-flip kinetics was taken at $T=0$ alternatingly with random
nearest-neighbour spin-exchanges ( Kawasaki-exchange at $T=\infty$).
The mean-field limit of this model together with results of the generalised
mean-field theory (GMF) have been presented in \cite{meor}.

In the present paper we further investigate 
NEKIM but now from the point of view of the underlying 1D spin
system with the aim of getting some more insight into the nature
of the  PC transition. 
Results of computer simulations are presented for different
critical exponents connected with the structure factor. The new
computational technique based on short-time dynamics and finite
size scaling, as introduced
and applied by Li,Sch\"ulke and Zheng \cite{Li1,Li2} for calculating
moments of the magnetisation and the time dependent Binder cumulant,
is also applied  here besides the usual finite size scaling (FSS)
 and time dependent simulations. In all our  numerical studies the
initial state is random with zero magnetisation and concerning
finite size effects the use of
antiperiodic boundary
conditions instead of the usual periodic ones will prove to be
essential.

In the field of domain-growth kinetics it has long been  accepted that
the scaling exponent of $L(t)$, the characteristic domain size,
is equal to $1/2$ if the order parameter is non-conserving.
Now let us restrict ourselves to a 1D Ising spin chain of length $L$
and define
the structure factor as usual :
$S(0,t)=L[<M^2>-<M>^2],\,\, M=\frac{1}{L}\sum_{i}s_i$,  ($s_i=\pm1$).
If the conditions of validity of scaling are fulfilled \cite{sad}
then
\be
S(0,t))\propto[L(t)]^d;\qquad L(t)\propto t^x
\ee
where $d=1$ now and $x=1/2$.
An other usually considered quantity is the excess energy of a monodomain
sample at the temperature of quench, which in our case is
proportional to the kink density
$n(t)=\frac{1}{L}<\sum_{i}\frac{1}{2}(1-s_{1}s_{i+1})>$.
\be
n(t)\propto\frac{1}{L(t)}\propto t^{-y}
\label{intro}
\ee
with $y=1/2$ in the Glauber-Ising case, expressing the well-known 
dependence on time of
annihilating random walk.

It is also well-known, that  the parameter  $p_T=e^{-4J/kT}$
 can be regarded as the quantity measuring the deviation from the
critical temperature $T=0$ of the 1D Ising model, and the usual (equilibrium)
critical exponents can be defined as powers of $p_T$. From exact solution 
the critical exponents of the spin-
susceptibility, coherence length and magnetisation are known to be
$\gamma=\nu=1/2$, $\beta=0$ ,
respectively. 
  Fisher's static scaling law $\gamma=d\nu-
2\beta$ is valid. Moreover, the dynamic critical exponent $Z$ is equal
to $2$.                

Instead of Glauber kinetics, let us apply now the nonequilibrium
NEKIM kinetics, and investigate the Ising system at and in the
immediate neighbourhood of one of its PC points. 
 On the level of kinks the PC point, separating the active and
absorbing phases, is a second order transition point
, while from the point of view of spins the absorbing phase
consists of a multitude of 1D Ising critical points, which ends
at the PC point.
We are interested in the behaviour of the spin system at this 
end-point.

The result is the following: the critical kink-dynamics has strong influence
on the spin-kinetics and even on its statics. Domain growth is governed
by criticality: x=1/Z, but the  
 dynamic exponent $Z$ changes
from $Z=2$ to $Z=1.75$. As to statics,  $\gamma=\nu=.444$ while  
$\beta=0$ and Fisher's
scaling law remain valid.

More spectacular is the change of the above $y$-exponent: $y=x/2 ; x=.57$ and
similarly the exponent $\alpha_n$ of the finite-temperature kink-density:
$\lim_{t\to\infty}n(t,p_T)\propto {p_T}^{\alpha_n}$,
which is $1/2$ in the Glauber-Ising limit, decreases to $\alpha_n=
\gamma/2=.222$.  We shall argue that this factor of $2$ between magnetic-
and kink-exponents 
has its origin in the
hyperscaling law, introduced for DP by Grassberger and
de la Torre \cite{grat}, which connects exponents of time-dependent
kink density and cluster-size.

These numerical results will be reported in detail in the following.
Moreover, we shall present further results, mainly from the side of 
 applying GMF approximations, 
by introducing a magnetic field-term into the spin-flip
probability, which causes the PC transition to become DP-like \cite{parkh}.
\vglue 0.5cm

\section{ The model }
\medskip

The model we will investigate here is a one-dimensional kinetic Ising
model evolving by a combined spin-flip and spin-exchange dynamics 
as described in \cite{men94}.
The
spin-flip transition rate 
in one-dimension
for spin $s_i$
sitting at site $i$  is:
\be
w_i = {\frac{\Gamma}{2}}(1+\delta s_{i-1}s_{i+1})\left(1 - 
{\gamma\over2}s_i(s_{i-1} + s_{i+1})\right), 
\ee
where $\gamma=\tanh{{2J}/{kT}}$ ($J$ denoting the coupling constant in
the Ising Hamiltonian), $\Gamma$ and $\delta$ are further
 parameters.
While in \cite{men94}  $T=0$ ($\gamma=1$)  has been taken,
we shall consider now finite temperature effects, too. Instead of
 $\gamma$ the
parameter $p_T$ will be used in the following.
The three independent rates:
\be
w_{indif}={\Gamma\over2}(1-\delta),\,\,
 w_{oppo}={\Gamma}(1+\delta){1\over{1+p_T}},\,\,
w_{same}={\Gamma}(1+\delta){{p_T}\over{1+p_T}},\,\,\,
 \ee
 where the subscripts of $w$ refer to the three possible
neighbourhoods of a given spin, are responsible -  on the level of 
domain walls -
 for random walk, annihilation and
pairwise  creation  (inside of a domain) of kinks, respectively.\\
The other ingredient of NEKIM has been
a spin-exchange transition rate of neighbouring spins (
the Kawasaki\cite{kaw72} rate at $T=\infty$):
\be
w_{ii+1}={1\over2}p_{ex}[1-s_is_{i+1}],
\ee
where $p_{ex}$ is the probability of spin exchange.

 Spin-flip and spin-exchange have been applied
alternatingly at each time step, the spin-flip part has been
applied using two-sublattice updating, while making $L$  
MC attempts at random ( $L$
denotes the size of the chain) has been counted as one time-step
of exchange updating.
In this system, at $T=0$, PC   type
phase transition takes place. In \cite{men94} we have started from a random
initial state and spotted the phase boundary in the ($\delta, pex$) plane.
In the following we will choose a typical point on this phase diagram
and make simulations at and around this point, fixing $\Gamma$ and
$p_{ex}$ and changing only $\delta$.
 The parameters chosen are:
$\Gamma=.35, pex=.3, \delta_c=-.395(2)$. We note here that $\Gamma$ appeared
as $1/\Gamma$ in ref.\cite{men94} and $\delta_c$ has a more accurate value
now than previously.

\section {Scaling forms and laws}

In the following we will be interested in two quantities characterising
the behaviour of the NEKIM, namely the structure factor or
spin-susceptibility and the kink-density under the conditions of a quench.
Thus, in contrast to the usually considered evolution from a pair of kinks,
we will restrict ourselves to completely random initial states ( $T=\infty$,
$M(0)=0$) and follow the development of the system via the rules described
in the previous section.
With $p_T$ defined above and $\epsilon=\mid{\delta-\delta_c}\mid$,
where $\delta<0$ is the parameter which drives the phase transition
in the present case, the scaling forms for $S$ and $n$ are as follows:
\begin{equation}
S(p_T,\epsilon,L,t)=t^xf(\frac{t^{\frac{1}{Z_c}}}{\xi}\ ,\frac{t^{\frac{1}{Z}
}}{\xi_{\bot}},\frac{\xi}{L},\frac{\xi_\bot}{L})
\label{1.1}
\end{equation}
\begin{equation}
n(p_T,\epsilon, L, t)=t^{-y}g(\frac{t^{\frac{1}{Z_c}}}{\xi}\ ,\frac{t^{\frac{1}{Z}
}}{\xi_{\bot}},\frac{\xi}{L},\frac{\xi_\bot}{L})
\label{1.2}
\end{equation}
with
$\xi={p_T}^{-\nu},\, \xi_\bot=\epsilon^{-\nu_\bot}$.
The exponents connected with $\epsilon$ have been written using the notation
of directed percolation, while those related to the temperature factor $p_T$
are written in the notation of equilibrium Ising system.
$Z$ and $Z_c$ are the respective dynamic critical exponents and we allow 
for the
possibility that they differ (though this will turn out not to be the case).

\subsection{The large L case}

Let us first take the limit $L\to\infty$. Then the dependences on $\xi/L$
and ${\xi_\bot}/L$ can be neglected in eqs(\ref{1.1}),(\ref{1.2}).
 If furthermore $\xi\to\infty$, $\xi_\bot\to\infty$
 the forms valid at the critical
point $\epsilon=0$, $p_T=0$ are obtained: $S_{c}(t)\propto t^x$ and
$ n_c(t)\propto t^{-y} $.
 Fig.1. shows $S_c(t)\propto t^x$, with the result
$x=.570(1)$. The same statistics  has led now for the exponent
of the kink density  to the value $y=.285(1)$ (denoted $\alpha$ in [9]).
\begin{figure}
\vspace{4mm} 
\centerline{\epsfxsize=6cm
                   \epsfbox{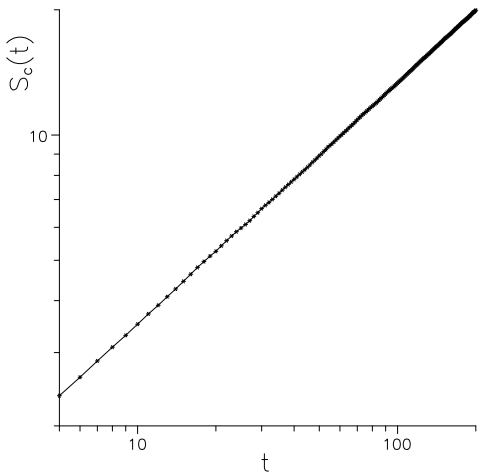}
                   \vspace*{4mm}      }
\caption{The sructure factor as a function of time at a typical
phase transition point of the NEKIM phase diagram.
$L=128$ and the number of independent initial  
configurations with zero magnetisation was  $10^5$. Here $t_{max}=200$
and finite size effects start to set in at about $t=1000$. The straight
line is fit with $x=.570$.}
\label{Fig1}
\end{figure}
 We have also made simulations
for the same quantities with
$L=8000$ up to $t=3\times 10^4$ and averaging over $2000$ initial states.
with similar result but with higher error and more than 100 times 
bigger computer
time.It is worth noticing, that within error $x=2y$.

Keeping $p_T=0$, near $\delta_c$ in the active phase eq.(\ref{1.2}) gives,
 for ${t\to\infty}$
the order parameter $n(\epsilon)\propto
\epsilon^{\beta_n}$ with the scaling law\cite{gra89}
\be
\beta_n=\nu_{\bot}Zy. 
\label{2.1.1}
\ee
A similar scaling relation can be obtained from (\ref{1.1}), which yields
for $t\to\infty$ : $ S(\epsilon)\propto \epsilon^{-\Theta}$
with
\be
\Theta={\nu_\bot}Zx.
\label{2.1.3}
\end{equation}
The divergence of the spin-susceptibility as a function of $\epsilon$
is understandable: in the subcritical regime (i.e. for $\mid \delta \mid
< \mid\delta_{c}\mid$
) it is infinite ( $T\rightarrow 0, t\rightarrow\infty$), because the whole
subcritical region is a plane of $1d$ critical (Ising) points.
In order to get $\Theta$ directly,
we have made simulations  for $S(\epsilon,t)$ around the chosen PC point 
in the interval 
$\epsilon=.02-.13$ 
 as shown on  Fig.2.
\begin{figure}
\vspace{4mm}
\centerline{\epsfxsize=6cm
                   \epsfbox{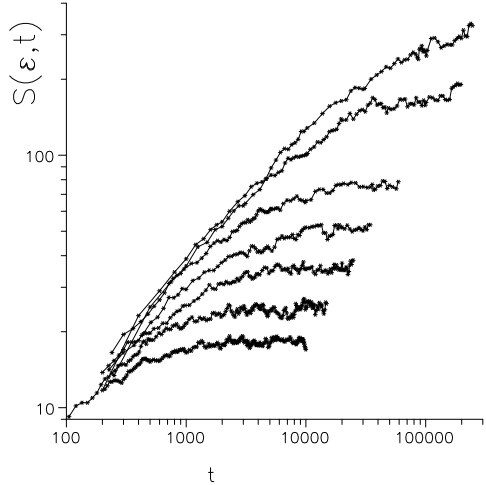}
                   \vspace*{4mm}  }  
\caption { The structure factor  $S(\epsilon,t)$ in the active
phase in the vicinity of the PC transition point for the following values
of $\epsilon$: $.13,\ .11,\ .09,\ .075,\ .06,\ .04,\ .03$ (from bottom
to top). L was varied between $2000$ and
$8000$ to avoid finite size effects before saturation sets in.
Number of independent states in averaging: typically $1000$.}
\label{Fig.2}
\end{figure}
The (time-averaged) saturation values of $S$ yield the exponent  $\Theta=
1.9(1)$ .
 From the same runs
as above we have obtained now $\beta_{n}=.88(4)$.
Thus within error $\Theta=2\beta_n$, which relation also follows
from eqs.\ (\ref{2.1.1}) and (\ref{2.1.3}) with $x=2y$. 
It is worth noting here that on the basis of the divergence of the spin 
susceptibility, $S(\epsilon)$, the PC transition point (as endpoint of
a line of first order transitions) 
can be found, without
any reference to kinks.

Taking now $\epsilon=0$ and keeping $p_T$ finite  in the limit  $t\to\infty$
we get
from eq.\ (\ref{1.2}) 
$n(p_T)\propto{p_T}^{\alpha_n}$ and the scaling relation
\begin{equation}
\alpha_n = y\nu Z_c.
\label{2.2.2}
\end{equation}
Fig.3 shows $n(t,p_T)$ as a function of $t$ at different values
of $p_T$ at $\epsilon=0$; the level-off values 
 could  be fitted with $\alpha_n=.222(5)$. This result is in accord
with the value for the $n=4$ BARW reported by Jensen \cite{jen94}
($1/{\delta_h}$ in his notation).

Similarly, taking eq.(\ref{1.1}) at $\epsilon=0$
and in the limit $t\to\infty$   the spin-susceptibility
arises: $ \chi\propto{p_T}^{-\gamma}$ together with the scaling relation
\begin{equation}
\gamma=x\nu Z_{c}.
\label{2.2.4}
\end{equation}
On Fig.4 $S(p_T,\epsilon=0,t)$ is plotted for different values of $p_T$.
\begin{figure}
\vspace{1mm}
\centerline{\epsfxsize=6cm
                   \epsfbox{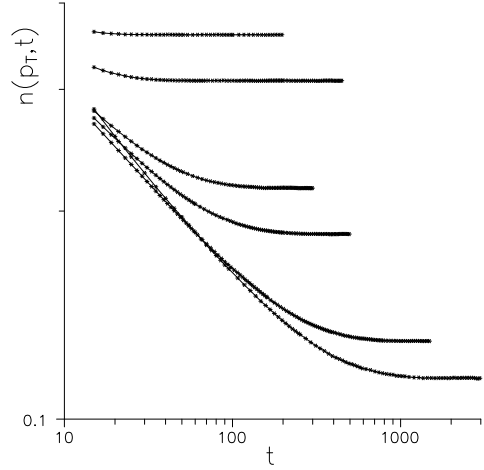}
                   \vspace*{4mm}  } 
\caption{The kink-density as a function of temperature and time at the
PC point.The curves are for $p_T=10^{-1},\ 5.0\times10^{-2},\ 10^{-2},\
5.0\times10^{-3},\ 10^{-3},\ 5.0\times10^{-4}$ from top to bottom.
 $L=2000$ and the number  of samples in averaging typically $2\times 10^4$.}
\label{Fig.3}
\end{figure}
\begin{figure}
\vspace{1mm}
\centerline{\epsfxsize=6cm
                   \epsfbox{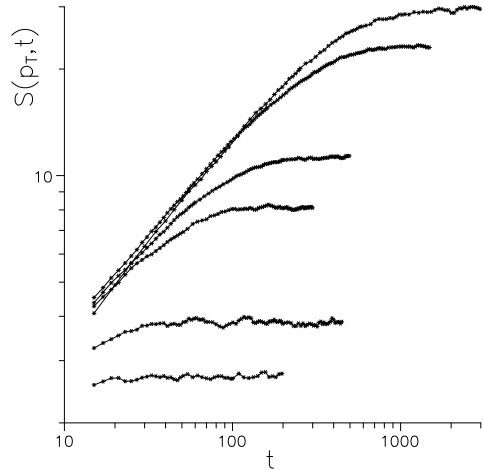}
                   \vspace*{4mm}  } 
\caption{The structure factor $S(p_T,t)$ for different values of $p_T$.
Details of simulation as of Fig.3  except that $p_T$
is decreasing from bottom to top here.}
\label{Fig.4}
\end{figure}

The temperature-dependent simulations
have been performed  in the
range $p_T=1.10^{-2} - 5.10^{-5}$. The level-off values  seen on
Fig.5, where the data for the two  highest temperatures ( $p_T=1.10^{-1},
 5.10^{-2}$) were discarded , yield the exponent of the spin-susceptibility 
at the
PC point as $\gamma=.445(5)$. This value is in accord, within error,
with the
scaling laws (\ref{2.2.2}) and (\ref{2.2.4}) which predict , using
the relation  $x=2y$ found above, $\gamma=2\alpha_n=.444$ .

\begin{figure}
\vspace{4mm}
\centerline{\epsfxsize=6cm
                   \epsfbox{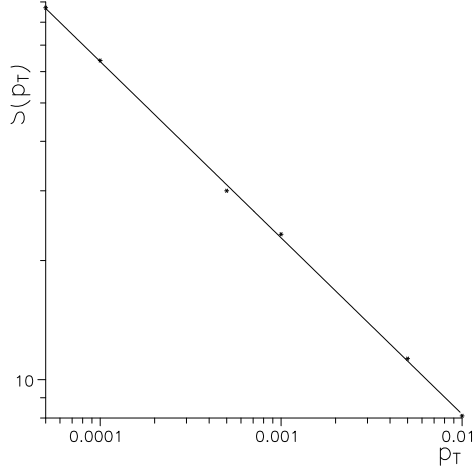}
                   \vspace*{4mm}  } 
\caption{ The saturation values of Fig.4 time-averaged, on a 
double-logarithmic scale.  Data for $p_T=10^{-4}$, $5\times 10^{-5}$, 
not shown on Fig.4, have been included. Straight line: fit of data
points with $\gamma=.445$}
\label{Fig.5}
\end{figure}

 It is to be noted, that due to the non-self-averaging
property of the structure factor  \cite{Heer}, all exponents
connected with this quantity have much larger statistical errors than
the ones connected with the kink density, and the
same applies to the time-dependent simulations. This explains
the large fluctuations exhibited on Fig.2 and  Fig.4 in comparison
with e.g. Fig.3.

\subsection{Static finite size scaling }

Finite size scaling will be used to find numerical values
for some more exponents.
The static FSS limit requires to take first the limit $t\to\infty$ 
and to suppose
$L\ll \xi, \xi_\bot$. Let us consider now the two possible orders of
limits to reach the PC point: a). $p_T=0$, $\epsilon\to 0$ and b).
$\epsilon=0$, $p_T\to 0$.

Case a). \\
Eqs (\ref{1.1}) and ({\ref{1.2}) lead to the expressions:
${\lim_{\epsilon\to 0}}S(\epsilon,L)
=L^{\frac{\Theta}{\nu_{\bot}}}f''(\frac{L}{\xi_\bot})
\propto L^{\frac{\Theta}{\nu_{\bot}}}$ and  \\ 
$\lim_{\epsilon\to 0}n(\epsilon,L)=L^{\frac{-\beta_{n}}{\nu_{\bot}}}
g''(\frac{L}{\xi_\bot})\propto L^{\frac{-\beta_{n}}{\nu_{\bot}}}$,
 respectively.

Case b).\\
From eqs (\ref{1.1}) and (\ref{1.2}) we get now
$\lim_{p_T\to0}S(p_T,L)=L^{\frac{\gamma}{\nu}}\tilde f''(\frac{L}{\xi})
\propto{L^{\frac{\gamma}{\nu}}}$ and \\
$\lim_{p_T\to0}n(p_T,L)=L^{-\frac{\alpha_{n}}{\nu}}\tilde g''(\frac{L}{\xi})
\propto{L^{-\frac{\alpha_n}{\nu}}}$, respectively.

Within error we have not found any (numerical) evidence against 
the supposition 
that the order of 
 limits $\epsilon\to 0$ and $p_T\to 0$ were interchangeable, 
  the same fixed point seems to be reached in both cases.
 ( Actually
this fact has expressed itself already in our finding, according to which
 the time-exponents
$x$ and $y$ are also the same independently from where we approach the
 limit
$\epsilon=0$, $p_T=0$ ; the difference should have been seen in
the preasymptotic time-dependence. E.g. for very small values of $p_T$
at $\epsilon=0$
$x=.57$, $y=.28$  was always clearly seen in the early-time behaviour
of $S(t)$ and $n(t)$, resp.).

Thus the above relations lead to the following scaling equalities:
\begin{equation}
\frac{\gamma}{\nu}=\frac{\Theta}{\nu_{\bot}},\qquad
\frac{\beta_n}{\nu_\bot}=\frac{\alpha_n}{\nu}
\label{3.1.14}
\end{equation}
In  numerical simulations with random initial states
usually periodic boundary conditions (pbc) are supposed. The fact that
pbc allow only an even number of 
kinks does not matter except under the conditions of FSS close to the
PC point  as eventually for all samples the ordering becomes perfect  
(depletion of kinks) and trivially
$\frac{\gamma}{\nu}=\frac{\Theta}{\nu_\bot}=1$. 
Because of the same reason it is not possible  to find the $L$-dependence
of the kink density, which breaks down as a function of time.
The proper procedure is to use
  antiperiodic boundary
conditions(apbc), which choice allows only  an odd number of kinks 
(i.e. all samples
are surviving)
provided the updating procedure is carefully done.
 As in the course of dynamic FSS one reaches  fixed points, which coincide 
with the
limiting values for $t\rightarrow\infty$ of $S$ and $n$ above, we will 
cite  our results 
in the next section.

As we shall see
 $\frac{\gamma}{\nu}=1$ is  valid also in case of antiperiodic boundary
conditions, which value together with
 eq.(\ref{2.2.4}) gives $Z_c=1/x$. Moreover, we shall arrive at $\frac{\beta_n}
{\nu_\bot}=1/2$, which, using eqs. (8) and (9), gives 
 $\frac{\Theta}{\nu_\bot}=1$ and leads to $Z=1/x$. Thus
 $Z_c=Z$, the two critical dynamical exponents coincide, as
anticipated.  It is worth noting that $Z=1/x$-type relation
 between dynamic critical exponent
and domain growth exponent has been found earlier for various $1d$ 
equilibrium
Ising systems  \cite{men90}.

\subsection{Early-time dynamic Monte-Carlo method}

In case of systems quenched to  their critical temperature \cite{Bray}
universality and scaling may appear in a quite early stage of time
evolution, far from equilibrium, where $\xi$ is still small. Based
on the scaling relation for such early time intervals, a new way for
measuring static and dynamic exponents has been proposed \cite{Li1,Li2}.
Now we apply  this method to get critical exponents for a nonequilibrium 
phase
transition.

Following \cite{Li1,Li2} we shall suppose the following relation to hold for
 the k-th moment
of the magnetisation near the critical point of the 1D spin system:
\be
M^{(k)}(t,p_T,L)=b^{-\frac{k\beta}{\nu}}M^{(k)}(b^{-Z}t,b^{1/\nu}p_T,b^{-1}L)
\label{3.2.1}
\ee
where  zero initial magnetisation has been considered and $b$ is a rescaling
factor ($b=2$ will be chosen).
 After generating randomly
an initial configuration, the system is let to evolve 
according to the nonequilibrium
kinetic rule of Section II at $p_T=0$. 
( We could have included $\epsilon$ in eq.(\ref{3.2.1})
as well, but  we will restrict ourselves to the case $\epsilon=0$,
$p_T=0$, so it is of no importance here). Average is taken over 
the initial configurations 
with zero
magnetisation. 
In order to get sufficiently good statistics  averaging has to be performed
over very many ($10^5 - 10^6$) independent initial states, as emphasised 
by  Li et al. \cite{Li1,Li2} who applied the method to the 2d Ising
model. To get the dynamical exponent $Z$ with great accuracy, they proposed
calculation of the time-dependent Binder cumulant:
\be
U(t,p_T,L)=1-\frac{M^{(4)}}{3{(M^{(2)})}^2}
\label{3.2.2}
\ee
which behaves at $p_T=0$ as
\be
U(t,0,L)=U(b^{-Z}t,0,L/b)
\label{3.2.3}
\ee
Similarly to eq.\ (\ref{3.2.1}), $n(t,L)$ can be supposed to behave according
to the relation:
\be
n(t,L)=b^{-\frac{\beta_n}{\nu_\bot}}n(b^{-Z}t,b^{-1}L)
\label{3.2.4}
\ee
Besides the triviality of the fixed points reached ($M^{(2)*}=1$,
$U^*=2/3$) the pbc case still leads - by proper fitting - to the value of
$Z$ and $\beta$. 
In case of $M^{(2)}$, using eq. (\ref{3.2.1}), 
collapsing of curves for different
values of $L$ has lead to $\beta=0.00(2)$, $Z=1.75(2)$; see
Fig.6.
In case of $U(t,0,L)$, which contains only $Z$ as  fitting parameter and thus
provides it directly, collapsing of curves could be achieved  again with 
$Z=1.75(1)$. Typically $10^5 - 3.10^5$ averages have been performed. 
\begin{figure}
\vspace{4mm}
\centerline{\epsfxsize=6cm
                   \epsfbox{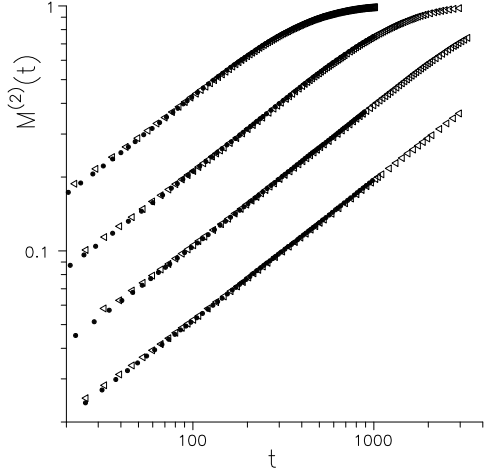}
                   \vspace*{4mm}  } 
\caption{ Collapsing of curves for $M^{(2)}$ with periodic boundary conditions
on a double-logarithmic scale.  $\triangle$ -s are rescaled data 
 while $\bullet$ -s
are original ones. From top to bottom: L=16($\triangle$)-32($\bullet$),
\, 32($\triangle$)-64($\bullet$),\, 64($\triangle$)-128($\bullet$),
\,128($\triangle$)-256($\bullet$). 
Number of samples with random initial states in
the averaging was typically $10^5$}.
\label{Fig.6}
\end{figure}

 With {\it antiperiodic boundary conditions} collapsing of curves has led us 
to the same results as above
concerning the exponents $\beta$ and $Z$. Under these conditions, however,
 important additional 
information  arises from 
 best fit to time averaged
saturation values.
 Concerning  
 $n(t,L)$, eq.(\ref{3.2.4}), the collapse of curves is illustrated with
  $L=100,200$ on Fig.7.
\begin{figure}
\vspace{1mm}
\centerline{\epsfxsize=6cm
                   \epsfbox{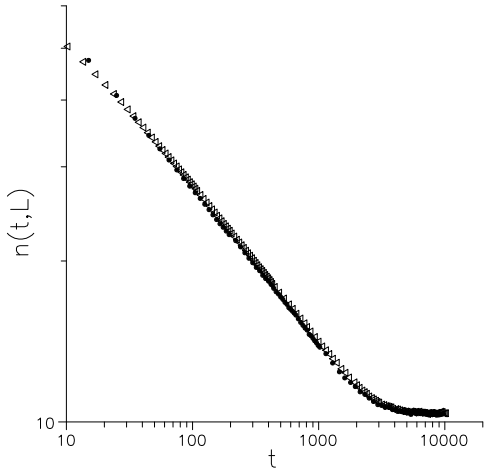}
                   \vspace*{1mm}  }
\caption{Collapsing of curves for $n(t,L)$ according to eq.(16). $\triangle$-s:
L=100, rescaled; $\bullet$-s: L=200. Number of samples in the averaging:
$10^5$. Boundary condition: antiperiodic, scale: double-logarithmic.}
\label{Fig.7}
\end{figure}The curves scale together with $Z=1.75(1)$ and 
 $\frac{\beta_{n}}{\nu_\bot}=.52(2)$.
 On the other hand,
fitting the time-averaged saturation values for $L=50,64,100,128,200,256$
gives for the kink-density $n(L)=.66\times L^{-.48(2)}$. Thus we conclude
$\frac{\beta_{n}}{\nu_\bot}(=\frac{\alpha_n}{\nu})=.50(2)$. 
This is in accord with the result of Jensen for the $n=4$ BARW \cite{jen94}.
\begin{figure}
\centerline{\epsfxsize=6cm
                   \epsfbox{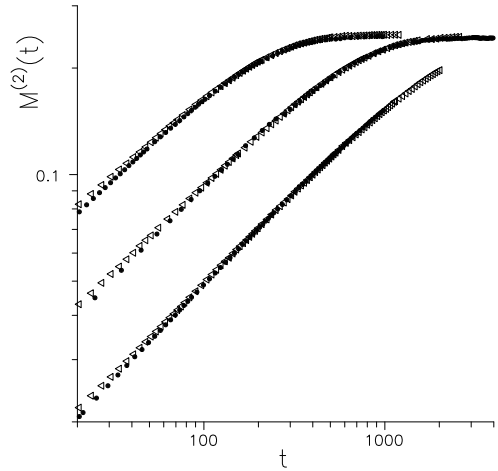}
                   \vspace*{1mm}  }
\caption{$M^{(2)}(t,L)$ with apbc. Rescaling with $b=2$, $\beta=0$ and
$Z=1.75$ has resulted in data points marked by $\triangle$ -s.
Collapsing of curves is shown -from top to bottom- for L=32($\triangle$)-
64($\bullet$),
\, 64($\triangle$)--128($\bullet$), \, 128($\triangle$)--256($\bullet$).
 Number of samples in averaging: $10^5 - 3\times 10^5$}
\label{Fig.8}
\end{figure}
\begin{figure}
\centerline{\epsfxsize=6cm
                   \epsfbox{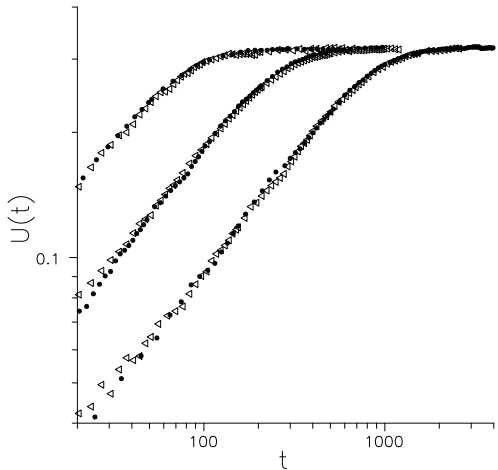}
                   \vspace*{1mm}  }
\caption{U(t,L) with apbc for L=16($\triangle$)-32($\bullet$),\,
32($\triangle$)-64($\bullet$),\, 64($\triangle$)-128($\bullet$) (from
top to bottom).
Best collapse of curves again wit $Z=1.75$. Number of samples in 
averaging : as for Fig.8.}
\label{fig.9}
\end{figure}

On Figs.8 and 9., respectively,
$M^{(2)}$ and $U$ are seen with antiperiodic boundary conditions for some
values of $L$. It is apparent, that $U$ is much more sensitive: here even
much more samples in averaging would have been needed to smooth out the curves.
The time-averaged saturation values for $L=50,64,100,128,200,256$ lead to
$S(L)=.26L^{.99(1)}$ (Fig.10), thus
$\frac{\gamma}{\nu}(=\frac{\Theta}{\nu_\bot})=.99(1) $.
\begin{figure}
\vspace{4mm}
\centerline{\epsfxsize=6cm
                   \epsfbox{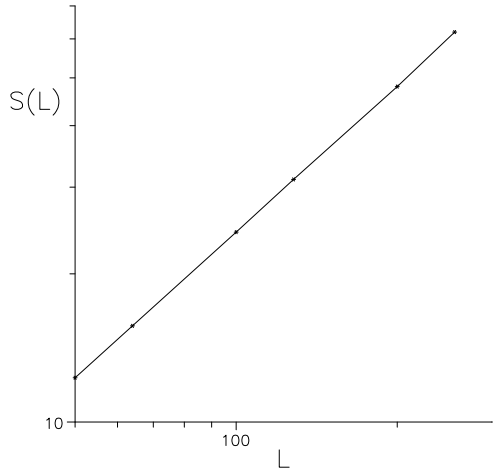}
                   \vspace*{4mm}  }
\caption{Time-averaged saturation values  $S(L)$ of $M^{(2)}(t,L)$ shown
on a double-logarithmic scale. Straight line: power-law fit with
${\gamma}/{\nu}=.99$. Averaging has been performed over  
$10^5 - 3\times 10^5$ independent random initial states with
apbc.}
\label{Fig.10}
\end{figure}
The difference as compared with the pbc case shows up in the prefactor 
( $.26(1)$
 instead of
 $1.0$ ).
For the Binder cumulant  $U^*=.32(1)$, again differing from the pbc value
of $2/3$.
For the sake of comparison, we have carried out the  same kind of 
simulations 
 for the Ising-Glauber case ($p_{ex}=0, \delta=0$) with
apbc, and obtained
 $M^{(2)*}_{Gl}=.333(3)$,
${n^*}_{Gl}=1.00(5)/L$, ${U^*}_{Gl}=.40(1)$ .
Here 
  $Z=2$ and $\beta=0$ have given the best fit and similar
 result can be expected  in the whole absorbing phase.
It is worth mentioning here, that these fixed-point values can be
derived exactly for the Glauber case with the result:
 $M^{(2)*}_{Gl}=1/3$,
${n^*}_{Gl}=1/L$, ${U^*}_{Gl}=.4$ \cite{ra}.

\section{Symmetry-breaking field}

It is by now well established that the PC transition has non-DP 
critical exponents, 
because of the modulo 2 conservation law. Park and Park \cite{parkh}
have introduced symmetry breaking external field in case of the 
interacting monomer-dimer model and showed that the DP universality
class is recovered if  one of the absorbing states is singled out.
They also have mentioned of having similar, though - to 
our knowledge -  not yet published
data for NEKIM and Grassberger's automata.
We have investigated the effect of an  external magnetic field $H$ 
on the NEKIM
model with simulations and with the help of the Generalized Mean-Field (GMF) 
technique as well, both confirming the DP behaviour. 
The transition probabilities of NEKIM, as given in Sect.II., are modified
in the presence of en external magnetic field H as:
\begin{eqnarray}
w_{indif}^h = w_{indif}(1 - h s_i), \\
w_{oppo}^h = w_{oppo}(1 - h s_i), \\
h = th({H\over kT}).
\end{eqnarray}
 Here we shall restrict ourselves to $p_T=0$. Fig.11 shows the
phase diagram of NEKIM in the $(h,\delta)$ plane, starting at  the
reference PC point for $h=0$ used in this paper ($\delta_c=-.395$).
\begin{figure}
\vspace{4mm}
\centerline{\epsfxsize=6cm
                   \epsfbox{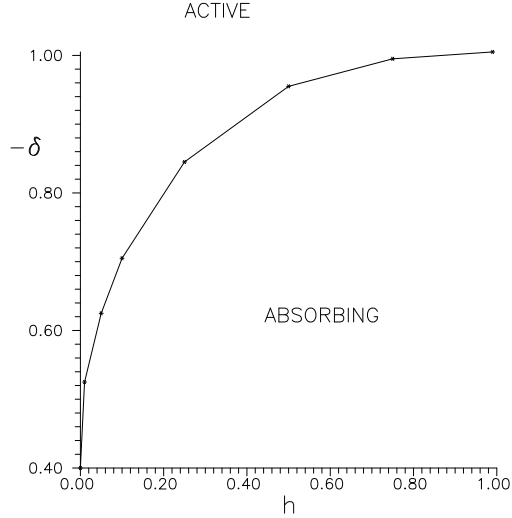}
                   \vspace*{4mm}  }
\caption{Phase diagram of NEKIM  in the $(h,\delta)$ plane 
in the presence of an external magnetic 
field.
 Parameters of the
transition probabilities: $\Gamma=.35, p_{ex}=.3$, see Section II.
Naturally, the phase diagram can be drawn symmetrically for negative
values of  $h$, as well.}
\label{Fig.11}
\end{figure}
We have applied only random
initial state simulations to find the points of the
line of phase transitions (critical exponents:  y=.17(2), $\beta_n=.26(2)$). 
It is seen, that with increasing field strength the critical point 
is shifted to more and more negative
values of $\delta$.

As to the treatment of the model with the use of the GMF technique, the
 details of which have been  explained in 
\cite{gut87,dic88,sza91}, in the field-free case  we have already obtained
estimates for the critical point and the effect of particle exchange earlier
 \cite{meor}. By applying the Coherent Anomaly 
extrapolation method (CAM) \cite{suz86} one can extrapolate  the 
critical exponents of the true singular behavior and in \cite{meor} we could 
give a rough 
estimate for the exponent $\beta_n$  of the PC transition 
based on $N\leq 6$-th 
order cluster GMF calculation. Now we extend the method for the 
determination of the exponent of  the order-parameter fluctuation as well :
\begin{equation}
\chi_n(\epsilon) = L(<n^2> - <n>^2) \sim \epsilon^{-\gamma_n} \ \ .
\end{equation}

The GMF equations have been set up for the steady states of NEKIM
in the presence of the H-field. The $N$-block
probabilities were determined as the numerical solution of the GMF
equations for $N = 1,...,6$. The traditional mean-field solution ($N = 1$)
results in stable solutions for the magnetization :
\begin{eqnarray}
M  = & - {h\over\delta}, \ \ & if \ \ \delta < 0 \ \ {\rm and} \ \ {h^2/
{\delta^2}}<1 \\
M  = & {\rm sgn}(h), \            & \  \  \  {\rm otherwise.}
\end{eqnarray}
and for the kink-concentration :
\begin{eqnarray}
n  = & {1\over 2}(1 - ({h\over\delta})^2 ), \ \ & if\ \  \delta<0 \ \ 
{\rm and} \ \  
{h^2/{\delta^2}}<1 \\
n  = & 0, \                                       & {\rm otherwise}.
\end{eqnarray}

For $N > 1$ the solutions can be found numerically only. By increasing the
order of approximation the critical point estimates $\delta_c(N)$ shift to 
more negative values similarly to the $H=0$ case. 
The $\lim_{N\to\infty}\delta_c(h)$ values have been determined with quadratic 
extrapolation in case of $h=0.01, 0.05, 0.08, 0.1$.
The resulting curves for $n(\delta)$ and $\chi_n(\delta)$ are shown on 
Figures 12. and 13., respectively for the case of $h=.1$ in different
orders $N$ of the GMF approximation.
\begin{figure}
\vspace{4mm}
\centerline{\epsfxsize=8cm
                   \epsfbox{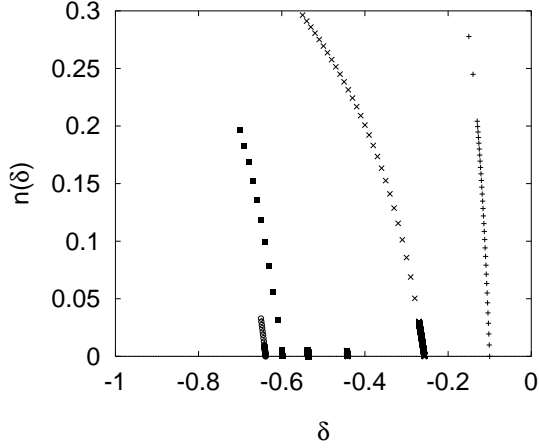}
                   \vspace*{4mm}  }
\caption{ The kink density in the neighbourhood of the critical
point $\delta_{c}(h)$ for $h=.1$. The curves from right to left correspond to
$N=1,...,6$ (level of GMF calculation). The points have been
determined with resolution of $10^{-5}$ in $\delta$, in order to
be able to extract CAM anomaly coefficients.}
\label{Fig.12}
\end{figure}
\begin{figure}
\vspace{4mm}
\centerline{\epsfxsize=8cm
                   \epsfbox{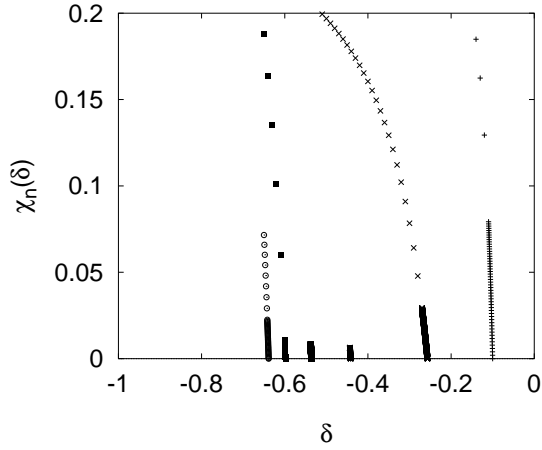}
                   \vspace*{4mm}  }
\caption{The same as Fig.12 but for the second moment
of the kink density.
 It is apparent also here  that the value
of $\delta_c(h)$, being equal to $.70(1)$ for $h=.1$ 
according to simulations, is fairly well approximated by the 
GMF-results for $N=5,6$.}
\label{Fig.13}
\end{figure}
Naturally, these curves exhibit mean-field type singularity at the critical
point :
\begin{eqnarray}
n(N) & \sim  \overline{\rho}_n (\delta/\delta_c(N)-1)^{\beta_{MF}} \\
\chi_n(N) & \sim  \overline{\chi}_n(N) (\delta/\delta_c(N)-1)^{-\gamma_{MF}},
\end{eqnarray}
with $\beta_{MF} = 1$ and $\gamma_{MF} = -1$.
According to the CAM (based on scaling) the critical exponents of 
the true singular behavior can be obtained via the scaling behavior
of anomaly factors:
\begin{eqnarray}
\overline{n}(N) & \sim  \Delta^{\beta_n - \beta_{MF}} \\
\overline{\chi_n}(N) & \sim  \Delta^{\gamma_n - \gamma_{MF}} \,
\end{eqnarray}
where we have used the $\Delta = 
(\delta_c / \delta_c(N) -  \delta_c(N) / \delta_c)$ 
invariant variable instead of $\epsilon$, that was introduced to 
make the CAM results independent of using $\delta_c$ or $1 / \delta_c$ coupling
(\cite{kol94}).
Since the level of the GMF calculation, what we could solve is $N \leq 6$,
we have taken into account correction to scaling, and determined the 
true exponents with non-linear fitting form :
\begin{eqnarray}
\overline{n}_(N) = a \ \Delta^{\beta_n - \beta_{MF}} 
+ b \ \Delta^{\beta_n - \beta_{MF} + 1} \\
\overline{\chi_n}(N) = a \ \Delta^{\gamma_n - \gamma_{MF}} 
+ b \ \Delta^{\gamma_n - \gamma_{MF} + 1} \\
\end{eqnarray}
where $a$ and $b$ are coefficients to be varied.
The results for various $h$-s are shown in table below.
For $h=0.0$ we could not determine the $\gamma_n$ exponent, 
because the low level GMF calculations resulted in discontinuous 
phase transition solutions -- what we can not use in the CAM 
extrapolation --  and so we had too few data points to achieve a
stable non-linear fitting. Higher order GMF solutions would help,
but that requires the solution of a non-linear set of equations
with more than $72$ independent variables. 
This problem does not occur  for $h \ne 0$; the above results - being 
based on all  $N=1..6$ point approximations - are fairly stable.
\begin{table}
\caption{CAM calculation results }
\begin{tabular}{lrrrrrl}
h       & 0.0 & 0.01 & 0.05 & 0.08 &  0.1 & DP \\
\tableline
$\beta_n$ & 1.0 & 0.281& 0.270& 0.258& 0.285& 0.2767(4) \\
$\gamma_n$&  -- & 0.674& 0.428& 0.622& 0.551& 0.5438(13) \\
\end{tabular}
\label{tablex}
\end{table} 

\section{Summary and Conclusions}

Time dependent simulations, FSS and  dynamic early-time  MC method 
have been applied here to investigate the
behaviour at and in the vicinity of a PC transition of the 
 nonequilibrium kinetic Ising model with the aim to complete earlier
results. Emphasis has been put now on the critical properties of
the 1d spin system underlying kinks.
In this way we have arrived at a more-or-less complete picture of the
effect the PC transition exerts on the statics and dynamics of the
1D Ising spin phase transition: we have found $Z=1.75$ (instead of
$Z=2.0$); $\gamma=\nu=.444$ (instead of 1/2). The PC point is endpoint
of a line of first order phase transitions ( by keeping
$p_{ex}$ and $\Gamma$ fixed and changing $\delta$ through negative values
to $\delta_c$); , where still $\beta=0$ holds
and also Fisher's scaling law: $\gamma=d\nu-2\beta$ is valid.
The PC point  has been approached from two directions: from the
active phase by changing $\epsilon=\delta_c -\delta$ and from the
direction of finite temperatures by  varying  $p_T=e^{-4J/kT}$.
The second moment of the magnetization ( structure factor ) provides
the 'magnetic' exponents 
$x, \Theta, \gamma$, which have been found - within error - 
to be twice as big as the
corresponding kink exponents, $y, \beta_n, \alpha_n$. The cause
of this factor of two must lie in the nature of the active phase,
of course, and it is
sufficient to understand $x=2y$ (time-dependent exponents at PC) as
 the rest follows from  scaling relations.

The idea is to recognize that there are two characteristic 
{\it growth} lengths at PC which have to have the same critical exponents.
Namely,  the magnetic one $L(t)\propto t^x$ (see Introduction)
and the cluster size defined through the square-root of 
$<R^{2}(t)>\propto t^z$. 
The latter one is obtained by starting either from two neighbouring kink 
initial states (see e.g.\cite{jen94}), or from a single kink 
\cite{gra89,men94}, while
 the magnetic domains grow in the quenching situation i.e.
from random initial states ($T=\infty\rightarrow T=0$). 
Both length-exponents
are, however connected with $Z$, the dynamic critical exponent, since 
 at PC  the only dominant length is the ( time-
dependent) correlation length.  $\frac{z}{2}=\frac{1}{Z}$ has been
shown to  follow from scaling in \cite{gra89} and \cite{mend} for one-kink
 and  two-kink initial states, respectively.
Presently  we have found  $x=1/Z$, thus  $x=z/2$ follows.

Exponents of kink density and cluster growth are connected by a
hyperscaling relation first established by Grassberger and de la Torre
\cite{grat} for the directed percolation transition. 
In the same form it does not apply to PC transition \cite{jen94,jen92},
where dependence on the initial state (one or two kinks) manifests
itself in two cluster-growth quantities: the kink-number $N(t)\propto
t^\eta$ and the survival probability $P(t)\propto t^{-\delta}$.  
(This $\delta$ has, of course, nothing to do
with the parameter $\delta$ of NEKIM)
 In case of odd number of
kinks (when $\delta=0$, as all samples survive) hyperscaling 
reads\cite{gra89}: $y+\eta=z/2$
As both $y$ and $\eta$ are exponents of time dependent kink density
at PC, they should agree numerically if scaling is valid with a single
(time-dependent) characteristic length, and this
is indeed the case (see e.g.\cite{jen94} for simulation data).
Consequently $2y=z/2=x$ follows from the hyperscaling relation. 

For  an even number of kinks (two-kink
situation) the relation with the random initial-state situation
investigated here can be established as follows. The local kink density
is defined as $\rho(x,t)\propto t^{\eta-z/2}f(x^2/{t^z})$  which
 divided by $P(t)$ leads to our $n(t)$ of eq.(\ref{intro}) at PC:
$n(t)\propto t^{-y}$ with $y=-\eta-\delta+z/2$. 
Using the hyperscaling relation valid in the two-kink case, namely
 $2\delta+\eta=z/2$,  $y=\delta$ follows. Thus the hyperscaling relation
can  be written as $2y+\eta=z/2$. 
Taking into account the result first obtained numerically by Jensen 
\cite{jen94}
that $\eta=0$ for the PC transition when starting with two kinks,
$2y=z/2=x$ holds, again. 

In this way the factor of two between
magnetic and kink exponents found in this paper could be explained
as following from scaling.

In the last chapter we have introduced a magnetic field into the
NEKIM transition rates and investigated its effect mainly in the
framework of generalized mean field approximation. By going up to
N=6 order cluster approximation, the expectation that the universality
class of the phase transition turns into DP-type \cite{parkh} has
been given support: we have found values for the exponents of the kink
density and its second moment which are very close to the corresponding
DP values.
\\
\\
{\bf ACKNOWLEDGEMENTS}\\
 The  authors would like to thank Z.R\'acz for numerous
 helpful remarks  and  the Hungarian research fund OTKA ( Nos.
T017493 and 4012) for
support during this study. The simulations were partially carried out on
the Fujitsu AP1000 parallel supercomputer.

\end{document}